\documentclass[preprint,12pt]{elsarticle}
\usepackage{graphicx}%
\usepackage{amsmath}
\usepackage{amssymb}
\usepackage{dcolumn}

\begin{document}

\begin{frontmatter}

\title{A class of solvable coupled nonlinear oscillators with amplitude independent frequencies}

\author{V.~K.~Chandrasekar\footnote{Corresponding author's e-mail : chandru25nld@gmail.com, phone/fax :~+91-431-2407093}}%

\author{Jane H.~Sheeba}%

\author{R. Gladwin Pradeep}%

\author{R.~S.~Divyasree\footnote{Department of Physics, Amrita University, Kollam, Kerala, India}}%

\author{and~ M.~Lakshmanan}%
\address{Centre for Nonlinear Dynamics, School of Physics,
Bharathidasan University, Tiruchirappalli - 620 024, Tamilnadu, India}

\date{\today}

\begin{abstract}
Existence of amplitude independent frequencies of oscillation  is an unusual property for a nonlinear oscillator. We find
 that a class of N coupled nonlinear Li\'enard type oscillators exhibit this interesting property. We show that a specific subset can be explicitly solved from which we demonstrate the existence of periodic and quasiperiodic solutions. Another set of $N$-coupled nonlinear oscillators, possessing the amplitude independent
 nature of frequencies, is almost integrable in the sense that the system can
 be reduced to a single nonautonomous first order scalar differential equation
 which can be easily integrated numerically
.
\end{abstract}


\begin{keyword} Nonlinear oscillators, coupled ordinary differential 
equations, complete integrability, isochronous systems
\end{keyword}

\end{frontmatter}

\section{Introduction}
\label{sec1}
Let us consider the following nonlinear oscillator described by the so called modified Emden equation with linear forcing term \cite{Chandrasekar:05},
\begin{eqnarray}
\label{mod01a}
 \ddot{x}+3x\dot{x}+x^{3}+\omega^2 x=0.
\end{eqnarray}
Here $\omega$ is a parameter.  Equation (1) can be considered as the cubic anharmonic oscillator with additional position dependent damping type nonlinear force $3x\dot{x}$.  This type of equation has been well studied in the literature.  For example, Eq. (1) with $\omega=0$ arises in a wide range of physical problems: it occurs in the study of equilibrium configurations of a spherical gas cloud acting under the mutual attraction of its molecules and subject to the laws of thermodynamics \cite{emden2,emden3}  and in the modelling of the fusion of pellets \cite{emden4}. It also governs spherically symmetric expansion or collapse of a relativistically gravitating mass \cite{emden5} . This equation can also be thought of as a one-dimensional analog of the boson ‘gauge-theory’ equations \cite{emden6,emden1}.

\begin{figure}
\begin{center}
\includegraphics[width=14cm]{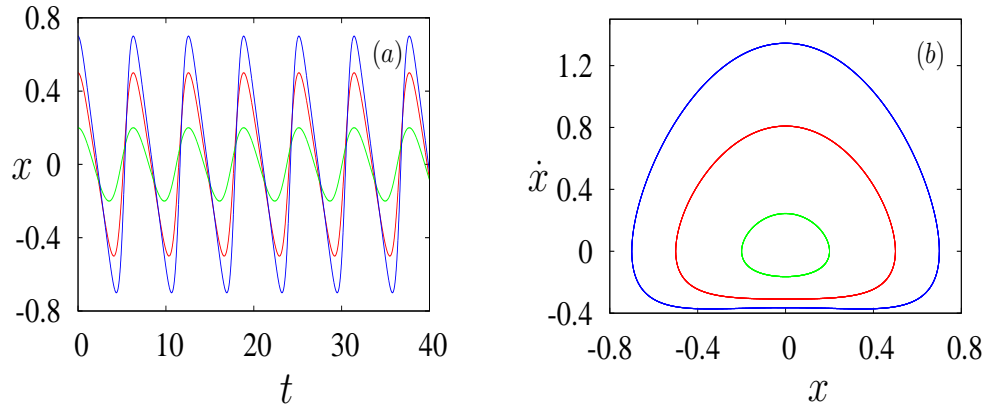}
\caption{(Color online) (a) Time series plot of Eq. (\ref{mod01a}) exhibiting periodic oscillation for three different initial conditions (three different colours) and $\omega=1.0$ (b) Phase space portrait of Eq. (\ref{mod01a})} 
\label{fig1}
\end{center}
\end{figure}

Equation (\ref{mod01a}) has been shown to posses an unusual property which is not a general characteristic of a nonlinear equation: The frequency of oscillation of the oscillator is independent of the amplitude similar to that of a linear harmonic oscillator \cite{Chandrasekar:05}. An oscillator which possesses this property is also known as an isochronous oscillator \cite{Calogero:11}. For a detailed study about isochronous orbits and isochronous oscillators one may refer to Refs. \cite{Calogero:11,Carinena:08}. Equation (\ref{mod01a}) admits the following nonsingular, periodic solution:
\begin{eqnarray}
\label{mod02a}
 x(t)=\frac{A\sin{(\omega t+\delta)}}{(1-\frac{A}{\omega}\cos(\omega t+\delta))}\,\,,\quad 0\le A<\omega.
\end{eqnarray}
Here $A$ and $\delta$ are arbitrary constants, expressible in terms of the two integrals of motion or integration constants obtained by solving (\ref{mod01a}) (for details see ref. \cite{Chandrasekar:05}). Note that the angular frequency of oscillation $\omega$ continues to be the same as that of the linear oscillation.  From this solution it is obvious that for $0<A<\omega$, equation (\ref{mod01a}) 
exhibits the property of amplitude independence of the frequency of oscillation. One can starightforwardly write down the solution of the initial value problem from the general
solution (\ref{mod02a}).  For example, for the initial condition $x(0)=B=\frac{A}{\sqrt{1-\frac{A^2}{\omega^2}}}$, $\dot{x}(0)=0$, from (\ref{mod02a}) we have the solution as
\begin{eqnarray}
x(t)=\frac{B\omega\sin\left[\omega t+\cos^{-1}\left(\frac{B}{\sqrt{B^2+\omega^2}}\right)\right]}{\sqrt{B^2+\omega^2}-B\cos\left[\omega t+\cos^{-1}\left(\frac{B}{\sqrt{B^2+\omega^2}}\right)\right]}.
\end{eqnarray}
Note that $B$ is the amplitude of oscillation.  Figure \ref{fig1} shows the periodic oscillations admitted by Eq. (\ref{mod01a}) for three different sets of initial conditions $x(0)$ and $\dot{x}(0)$ with $\omega=1.0$ in terms of three different colours.  We note here that the frequency of the oscillations is independent of the initial conditions as in the case of the linear harmonic oscillator.

One can trace the origin of this property of equation (\ref{mod01a}) to the fact that it can be transformed to the linear harmonic oscillator equation, 
\begin{eqnarray}
\ddot{U}+\omega^2 U=0,
\label {horm1}
\end{eqnarray}
through a nonlocal transformation,
\begin{eqnarray}
U=\displaystyle{x(t) e^{\int_0^t x(t')dt'}}.
\label {lam103}
\end{eqnarray}
The solution (\ref{mod02a}) can be obtained (see below, equation (\ref{nld05})) from the solution of (\ref{horm1}), $U=A\sin(\omega t+\delta)$,
where $A$ and $\delta$ are arbitrary constants and the frequency, 
$\omega$, is independent of the amplitude. Such a linearization property is one of the fascinating features associated with a class of nonlinear equations exhibiting large number of symmetries and extensive search for such linearizing transformations is being made in the recent literature \cite{Chandrasekar:05,Iacono:11,Gladwin:09,Chandrasekar:06}.
In fact, there exists a class of nonlinear oscillators
which are connected to the linear oscillator equation (\ref{horm1}) through the following nonlocal transformation \cite{Chandrasekar:06}
\begin{eqnarray}            
U=x(t)e^{\int_0^t{f(x(t'))}dt'},
\label {int02}
\end{eqnarray}
where $f(x(t))$ is an arbitrary function of $x(t)$. Now substituting  (\ref{int02})
into (\ref{horm1}) we get a nonlinear ordinary differential equation (ODE) of the form
\begin{eqnarray}            
\ddot{x}+(2f+xf')\dot{x}+(f^2+\omega^2)x=0,
\label {int03}
\end{eqnarray}
where prime denotes differentiation with respect to $x$. Equation (\ref{int03}) is a special case of the well known Lienard equation (LE) \cite{lienard}
\begin{eqnarray}
\label{mod01b}
 \ddot{x}+u(x)\dot{x}+v(x)=0.
\end{eqnarray}

One can also consider a more general nonlocal transformation of the form
\begin{eqnarray}
U=g(x(t))e^{\int f(x(t'))dt'},
\end{eqnarray}
and substituting this in (\ref{horm1}) we get
\begin{eqnarray}
\ddot{x}+\frac{g''}{g'}\dot{x}^2+\frac{gf'}{g'}\dot{x}+2f\dot{x}+\frac{\omega^2 g}{g'}+\frac{f^2g}{g'}=0, \quad \left(g'=\frac{dg}{dx}\right).
\end{eqnarray}
We find the above equation reduces to a Li\'enard type equation only for the choice $g(x)=x$.
Interestingly for $f=x^{-2}$, equation (\ref{int03}) becomes the well known  isotonic oscillator \cite{Carinena:08} equation,
\begin{eqnarray}
\label{mod01e}
 \ddot{x}+\omega^2 x+\frac{1}{x^3}=0.
\end{eqnarray}

The solution of the nonlinear equation (\ref{int03}) is obtained by using the identity 
\begin{eqnarray}            
\frac{\dot{U}}{U}=\frac{\dot{x}}{x}+f(x(t)).
\label {nld05}
\end{eqnarray}
Since $U=A\sin{(\omega t+\delta)}$, where $A$ and $\delta$ are integration constants, is the solution of the linear harmonic oscillator 
(\ref{horm1}), equation (\ref{nld05}) can be rewritten as the first order nonlinear differential equation of form
\begin{eqnarray}
\label{mod07aa}
\dot{x}+f(x(t))x-\frac{\omega x}{\tan(\omega t+\delta)}=0.
\end{eqnarray}
Now one can get the solution of (\ref{int03}) by solving (\ref{mod07aa}). In particular, for the specific case $f=x^q$ equation (\ref{mod07aa}) becomes a Bernoulli equation of the form
\begin{eqnarray}
\label{mod07b}
 \dot{x}=\frac{\omega x}{ \tan(\omega t+\delta)}-x^{q+1}.
\end{eqnarray}
The corresponding ODE (\ref{int03}) becomes
\begin{eqnarray}
\label{mod01}
 \ddot{x}+(q+2)x^{q}\dot{x}+x^{2q+1}+\omega^2 x=0,
\end{eqnarray}
and equation (\ref{mod01a}) is the special case corresponding to $q=1$.

\begin{figure}
\begin{center}
\includegraphics[width=14cm]{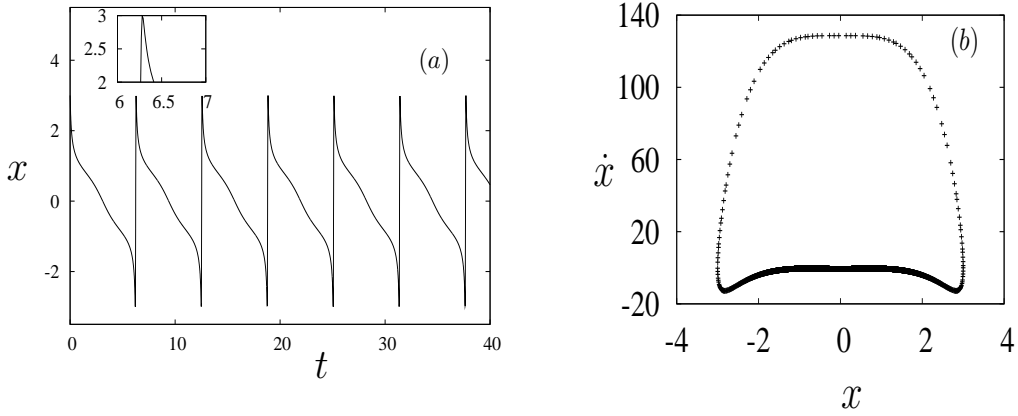}
\caption{(a) Time series plot of Eq. (\ref{mod01}) for $q=3$  exhibiting periodic oscillations with the initial condition $x(0)=3$ and $\dot{x}(0)=0$ for $\omega=1$ (b) Phase space portrait of Eq. (\ref{mod01})} 
\label{fig2}
\end{center}
\end{figure}

Upon integrating (\ref{mod07b}) we get the periodic solution of (\ref{mod01}) as
\begin{eqnarray}
\label{mod02}
&&\hspace{-1cm} x(t)=\frac{\sin{(\bar{t})}}{[I-\cos(\bar{t})(\sin^{2m}(\bar{t})+\sum_{k=0}^{m}C_{k}\sin^{2m-2k-2}(\bar{t})]^{\frac{1}{(2m+1)}}},
\end{eqnarray}
where $\bar{t}=\omega t+\delta$, $\vline\sum_{r=0}^m C_{k}\vline<I-1$, $q=2m+1,\,\;C_{k}=(2^{k+1}m(m-1)\ldots(m-k))/((2m-1)(2m-3)\ldots(2m-2k-1))$, $I$ and $\delta$ are arbitrary constants. Here $m$ is a non-negative integer and $\omega$ is the angular frequency. One can note that solution (\ref{mod02}) is also isochronous. This has indeed been reported recently by Iacono and Russo \cite{Iacono:11} using a different procedure.   In figure \ref{fig2} we show the periodicity of the solution for the case $q=3$ and with the initial conditions $x(0)=3$ and $\dot{x}(0)=0$.  
We additionally remark here that the case $q=2m$, $m=1,2,\ldots,$ of equation (\ref{mod01}) is also exactly solvable but the solutions are of damped oscillatory type as will be proved later in this article (Sec. \ref{sec2}) even for coupled systems.

In this paper we will show that this unusual (amplitude independent frequency) property possessed by the class of equations (\ref{int03}) is not limited to scalar LEs alone but is also possessed by a class of $N$ coupled nonlinear oscillator equations. In order to demonstrate the existence of periodic and quasiperiodic solutions for such systems, we extend the above mentioned procedure to $N$ coupled LEs. We also derive the integrals of motion and the general solution to the system of $N$ coupled LEs.  We also point out that there exists another class of $N$ coupled nonlinear oscillators which is almost integrable in the sense that it admits $(2N-1)$ independent integrals and reduces to a single first order nonautonomous nonlinear differential equation that can be solved numerically.  The frequency of oscillation of its periodic solutions is again independent of amplitude.

In the following section we show that a system of $N$ coupled LEs can be explicitly integrated to find periodic and quasiperiodic solutions whose frequencies are identical to a system of uncoupled linear oscillators. In Sec. \ref{sec3} we deduce the underlying $2N$ integrals of motion for the system and analyze their structure. In Sec. \ref{nearlyintegrable} we numerically solve the system of $N$ coupled LEs for which the explicit general solution is not known, but $(2N-1)$ integrals are known.
We finally summarize our results in Sec. \ref{sec4}.

\section{Coupled nonlinear oscillators of Lienard type and integrability}
\label{sec2}
Generalizing the above results of the scalar system (\ref{int03}) to a system of coupled nonlinear oscillators of Lienard type, we relate them to a system of uncoupled $N$-dimensional linear harmonic oscillators.
For this purpose, we consider the $N$-dimensional anisotropic  harmonic oscillator equation of the form
\begin{eqnarray}
\ddot{U}_i+\omega_i^2 U_i=0, \; i=1,2, \ldots, N, 
\label {nhorm1}
\end{eqnarray}
where the freqencies $\omega_i,\;i=1,2, \ldots, N$ are in general different.
Let us now introduce the nonlocal transformation
\begin{eqnarray}
U_i=\displaystyle{x_i e^{\int f_i(x_1, x_2, \ldots x_N)dt}=x_i e^{\int f_i(\bar{x})dt}}, \; i=1,2, \ldots, N,
\label {nhorm2}
\end{eqnarray}
where $f_i(\bar{x})=f_i(x_1, x_2, \ldots x_N)$ are arbitrary functions of the variables. Subsituting (\ref{nhorm2}) into (\ref{nhorm1}) we get a system of $N$ coupled Lienard type second order nonlinear oscillator equations of the form
\begin{eqnarray}
\label{nhorm3}
 \ddot{x}_i&+&\sum_{j=1}^{N} f_i^{(j)}(\bar{x})x_i\dot{x}_j+f_i(\bar{x})\dot{x}_i\nonumber\\
&&\qquad\qquad+f_i^2(\bar{x})x_i+\omega_i^2 x_i=0, \; i=1,2, \ldots, N,
\end{eqnarray}
where $f_i^{(j)}=df_i/dx_j$, $i,j=1,2, \ldots, N$. In the following we shall demonstrate the existence of periodic and quasiperiodic solutions with amplitude independent frequencies, $\omega_i,\;i=1,2, \ldots, N$ (isochronous property) for the above mentioned nonlinear oscillator equation with specific forms of $f_i$'s.

To obtain the desired results, we make use of the following identities,
\begin{eqnarray}            
\frac{\dot{U}_i}{U_i}=\frac{\dot{x}_i}{x_i}+f_i(\bar{x}), \; i=1,2, \ldots, N,
\label {nhorm4}
\end{eqnarray}
derived from equation~(\ref{nhorm2}). Obviously the solution of the system of second order linear ODEs 
(\ref{nhorm1}) is 
\begin{eqnarray}
U_i=A_i\sin{(\omega_it+\delta_i)}, \label{ham_sol}
\end{eqnarray}
where $A_i$'s and $\delta_i$'s, $i=1,2,\ldots, N$, are $2N$ integration constants. Now substituting the solution (\ref{ham_sol}) in  equation (\ref{nhorm4}) we find the following system of coupled first order ODEs to represent (\ref{nhorm3}),
\begin{eqnarray}
\label{mod07}
 \dot{x}_i=\frac{\omega_i x_i}{ \tan(\omega_it+\delta_i)}-f_i(\bar{x})x_i,\; i=1,2, \ldots, N.
\end{eqnarray}
It may be noted that $N$ integration constants, $\delta_i$'s, $i=1,2,\ldots, N$, of the $N$ coupled second order ODEs (\ref{nhorm3}) appear explicitly in (\ref{mod07}). In order to find the general solution of (\ref{nhorm3}) we need $N$ more integration constants which are to be obtained by integrating (\ref{mod07}). For general forms of $f_i(\bar{x})$ in (\ref{nhorm3}) this cannot be done. However for the special choice,
\begin{eqnarray}
f_k=f_N(\bar{x})=f(\bar{x})=f(x_1, x_2, \ldots x_N), \; k=1,2,\ldots, N-1, \label{mod07e}
\end{eqnarray}
equation (\ref{mod07}) can be integrated.  In order to perform this integration we multiply each equation of the system (\ref{mod07}) by $x_N$ for $i=1,2,\ldots,N-1$ and the last equation by $x_i$ and subtract them to get
\begin{eqnarray}
\dot{x}_kx_N-\dot{x}_Nx_k=\left(\frac{\omega_k}{\tan(\omega_kt+\delta_k)}-\frac{\omega_N}{\tan(\omega_Nt+\delta_N)}\right)x_kx_N,\quad k=1,2,\ldots,N.\label{int1}
\end{eqnarray}
Dividing Eq. (\ref{int1}) throughout by $x_kx_N$, we get
\begin{eqnarray}
\left(\frac{\dot{x}_k}{x_k}-\frac{\dot{x}_N}{x_N}\right)=\left(\frac{\omega_k}{\tan(\omega_kt+\delta_k)}-\frac{\omega_N}{\tan(\omega_Nt+\delta_N)}\right).
\end{eqnarray}
Integrating the above equation, we get
\begin{eqnarray}
\log x_k-\log x_N=\log \sin(\omega_kt+\delta)-\log \sin(\omega_Nt+\delta)+\log I_k.
\end{eqnarray}
Rewriting the above equation, we get
\begin{eqnarray}
\label{mod08}
 \frac{x_k}{x_N}=I_{k}\frac{\sin(\omega_kt+\delta_k)}{\sin(\omega_Nt+\delta_N)}=h_k(t), \;\;k=1,2,\ldots, N-1
\end{eqnarray}
where $I_{k}$'s, $k=1,2,\ldots, N-1$, are $N-1$ integration constants. We are also left with a first order ODE of the form  
\begin{eqnarray}
\label{mod07a}
 \dot{x}_N=\frac{\omega_N x_N}{\tan(\omega_Nt+\delta_N)}-f(\bar{h}(t),x_N)x_N,
\end{eqnarray}
where $\bar{h}(t)=\{h_1(t), h_2(t), \ldots, h_{N-1}(t)\}$ and $h_k(t)$'s are given by equation (\ref{mod08}). Now the problem of solving the set of $N$ coupled autonomous second order ODEs (\ref{nhorm3}) is reduced to the problem of solving a single non-autonomous first order ODE (\ref{mod07a}). Therefore one can get the general solution for equation (\ref{nhorm3}) for the case (\ref{mod07e}) by solving equation (\ref{mod07a}).

Equation (\ref{mod07a}) again cannot be in general solved explicitly for arbitrary form of the function $f$ \cite{murphy}. Hence, in order to solve (\ref{mod07a}) we assume that the function $f$ has a symmetry $f(\alpha\bar{x})=\alpha^q f(\bar{x})$, where $\alpha\bar{x}=\{\alpha x_1, \alpha x_2, \ldots \alpha x_N\}$, $\alpha$ and $q $ are arbitrary parameters. This implies that $f$ is a homogeneous polynomial and we assume the following form of $f$,
\begin{eqnarray}
\label{mod10e}
f=\sum_{l=1}^{N} \gamma_l(t)x_l^{q},
\end{eqnarray}
where $\gamma_l(t)$'s are arbitrary functions of $t$.  However, in the present case we assume $\gamma_l$'s to be constants only for simplicity.  Even when they are functions of $t$ the following integration procedure holds good.  With the above choice of $f$, equation (\ref{nhorm3}) reduces to the system of coupled nonlinear oscillator equations,
\begin{eqnarray}
\label{nhorm3a}
 \ddot{x}_i&+&2\sum_{l=1}^N \gamma_lx_l^{q}\dot{x}_i+q\sum_{l=1}^N \gamma_l x_i x_l^{q}\dot{x}_l
\nonumber\\&&\qquad+(\sum_{l=1}^N \gamma_lx_l^{q})^2x_i+\omega_i^2 x_i=0, \; \; i=1,2, \ldots, N.
\end{eqnarray}
The solution of this system of coupled nonlinear oscillators can be obtained by solving the following first order nonlinear differential equation obtained by substituting (\ref{mod10e}) into (\ref{mod07a}) along with (\ref{mod08}):
\begin{eqnarray}
\dot{x}_N=\omega_N\cot(\omega_Nt+\delta_N)x_N-x_N^{q+1}\sum_{l=1}^N\bar{\gamma}_l\frac{\sin^q(\omega_lt+\delta_l)}{\sin^q(\omega_Nt+\delta_N)},\label{bern-eq1}
\end{eqnarray}
where  $\bar{\gamma_l}=I_{l}\gamma_l$, $l=1,2,\ldots, N-1$, and $\bar{\gamma_N}=\gamma_N$ .
The above equation is of the first order Bernoulli equation type \cite{lienard,murphy}, namely
\begin{eqnarray}
\frac{dv}{du}=P(u)v+Q(u)v^n,
\end{eqnarray}
where $P(u)$ and $Q(u)$ are arbitrary functions of the independent variable $u$.
With the substitution $y(t)=x_N(t)^q$ in (\ref{bern-eq1}) we get the following first order linear inhomogenous ODE
\begin{eqnarray}
\frac{dy}{dt}=-q\omega_N\cot(\omega_Nt+\delta_N)y+q\sum_{l=1}^N\bar{\gamma}_l\frac{\sin^q(\omega_lt+\delta_l)}{\sin^q(\omega_Nt+\delta_N)}.\label{bern1}
\end{eqnarray}
The general solution of (\ref{bern1}) is obviously
\begin{eqnarray}
 y(t)&=&\frac{\sin^q(\omega_Nt+\delta_N)}{[I_N+q\int \sum_{l=1}^{N} \bar{\gamma_l} \sin^{q}{(\omega_lt+\delta_l)}]},
\end{eqnarray}
where $I_N$ is the integration constant.   Rewriting the above in terms of $x_N$, we get
\begin{eqnarray}
\label{mod11}
 x_N(t)&=&\frac{\sin{(\omega_Nt+\delta_N)}}{[I_N+q\int \sum_{l=1}^{N} \bar{\gamma_l} \sin^{q}{(\omega_lt+\delta_l)}]^{1/q}},
\end{eqnarray}
where $I_N$ is the $2N^{\mbox{th}}$ integration constant which we are looking for.

The integral appearing in the denominator of the above expression can be integrated explicitly for arbitrary values of $q$.  However, we find that the system admits oscillatory solutions only when $q$ is a positive integer and for this choice we find  the above expression reduces to either of the following forms depending on the value of $q$ \cite{Beyer:87}.

\noindent{\bf (i) Case 1 - $q$ odd :} We take $q=2m+1$, $m=0,1,2,\ldots$.   Then equation (\ref{mod11}) reduces to the form
\begin{subequations}
\label{sol_odd}
\begin{eqnarray}
\label{mod12}
&&\hspace{-1cm} x_N(t)=\frac{\sin\bar{t}_N}{[I_N-\sum_{l=1}^{N} \frac{\bar{\gamma_l}}{\omega_l}\cos{\bar{t}_l}(\sin^{2m}\bar{t}_l+\sum_{k=0}^{m-1}C_{k}^{o}\sin^{(2m-2k-2)}{\bar{t}_l})]^{\frac{1}{2m+1}}},
\end{eqnarray}
where $\bar{t}_l=\omega_lt+\delta_l$, $l=1,\,2,\,\ldots,\,N$ and $C_{k}^{o}=(2^{k+1}m(m-1)\ldots(m-k))/((2m-1)(2m-3)\ldots(2m-2k-1))$. 
Using  (\ref{mod08}) and  (\ref{mod12}) we get the remaining $N-1$ variables as 
\begin{eqnarray}
\label{mod13}
&& \hspace{-1cm}x_k(t)=\frac{I_{k}\sin{(\bar{t}_k)}}{[I_N-\sum_{l=1}^{N} \frac{\bar{\gamma_l}}{\omega_l}\cos{\bar{t}_l}(\sin^{2m}\bar{t}_l+\sum_{k=0}^{m-1}C_{k}^{o}\sin^{(2m-2k-2)}{\bar{t}_l})]^{\frac{1}{2m+1}}},
\end{eqnarray}
\end{subequations}
where $k=1,2,\ldots, N-1$.  Note that for the solution (\ref{sol_odd}) to be nonsingular periodic, we require the condition $(\bar{\gamma_1})/(\omega_l)\vline\sum_{k=1}^{m-1} C_{k}^o+1\vline<I_N$.

We note here that for the special case $m=0$ the solutions (\ref{mod12}) and (\ref{mod13}) become the periodic/quasiperiodic solution
\begin{eqnarray}
\label{mod12a}
 x_N(t)&=&\frac{\sin{(\bar{t}_N)}}{I_N-\sum_{l=1}^{N} \frac{\bar{\gamma_l}}{\omega_l}\cos{(\bar{t}_l)}},\nonumber\\
 x_k(t)&=&\frac{I_{k}\sin{(\bar{t}_k)}}{I_N-\sum_{l=1}^{N} \frac{\bar{\gamma_l}l}{\omega_l}
\cos{(\bar{t}_l)}}, \;\;k=1,2,\ldots, N-1.
\end{eqnarray}
which exactly matches with the solution given in \cite{Gladwin:09} for the so called coupled modified Emden equation.

\noindent{\bf (ii) Case 2 - $q$ even}, Here we take $q=2m$, $m=1,2,\ldots$.  Then equation (\ref{mod11}) reduces to the form
\begin{subequations}
\label{sol_even}
\begin{eqnarray}
\label{mod12b}
&&\hspace{-1cm} x_N(t)=\frac{\sin\bar{t}_N}{[I_N-D(t)+Bt]^{\frac{1}{2m+1}}},
\end{eqnarray}
where $D(t)=\sum_{l=1}^{N} \frac{\bar{\gamma_l}}{\omega_l}\cos{\bar{t}_l}(\sin^{2m-1}\bar{t}_l+\sum_{k=1}^{m-1}C_{k}^e\sin^{(2m-2k-1)}{\bar{t}_l})$, $B=((2m-1)!!)/(2^{m-1} (m-1)!)$,  $C_{k}^e=((2m-1)(2m-3)\ldots(2m-2k-1))/(2^k(m-1)(m-2)\ldots(m-k))$. Using (\ref{mod08}) and  (\ref{mod12b}) we get the remaining $N-1$ variables as
\begin{eqnarray}
\label{mod12c}
&&\hspace{-1cm} x_k(t)=\frac{I_k\sin\bar{t}_k}{[I_N-D(t)+Bt]^{\frac{1}{2m+1}}}\,\,.
\end{eqnarray}
\end{subequations}
From  (\ref{mod12b}) and (\ref{mod12c}) we find that equation (\ref{nhorm3a}) admits only oscillatory dissipative type solution for the choice $q=2m$, $m=1,2,\ldots$\,\,.


Note that for $q$ odd positive integer in (\ref{nhorm3a}), one can have either periodic or quasiperiodic solutions, depending on whether the uncoupled frequencies $\omega_i$'s are commensurate or not.   In Fig. 1 we have presented quasiperiodic and periodic solutions for suitable choices of the uncoupled frequencies $\omega_1,\,\omega_2,\,\ldots,\omega_N$ with $N=10$ in the form of projected phase space plots in the $x_1-x_3$ plane.  We find that for the choice $\omega_1=2$ and $\omega_2=\omega_3=\ldots=\omega_{10}=1$, the system (\ref{nhorm3a}) exhibits $2:1$ period oscillations.  Similarly for the choice $\omega_1=\sqrt{2}$ and $\omega_2=\omega_3=\ldots=\omega_{10}=1$, the system (\ref{nhorm3a}) is found to exhibit quasiperiodic oscillations.

One can note that equation (\ref{nhorm3a}) can also be rewritten in the following first order form as was done by Iacono and Russo \cite{Iacono:11} for the scalar case as
\begin{subequations}
\begin{eqnarray}
\label{mod05aa}
 \dot{x}_i&=&\omega_iy_i-f(\bar{x})x_i,\\ 
\label{mod05ab}
\dot{y}_i&=&-\omega_i x_i-f(\bar{x})y_i, \; i=1,2, \ldots, N.
\end{eqnarray}
\label {mod05}
\end{subequations}
Multiplying Eq. (\ref{mod05aa}) by $y_i$ and Eq. (\ref{mod05ab}) by $x_i$ and subtracting the resulting equations we get
\begin{eqnarray}
\dot{x}_iy_i-x_i\dot{y}_i=\omega_i(x_i^2+y_i^2).\nonumber
\end{eqnarray}
Dividing throughout by $y_i^2$ and rewriting we get
\begin{eqnarray}
\label{mod06}
 \frac{d}{dt}\bigg(\frac{x_i}{\omega_iy_i}\bigg)=1+\bigg(\frac{x_i}{y_i}\bigg)^2, \; i=1,2, \ldots, N.
\end{eqnarray}
Upon introducing the angle variable $\theta_i=\tan^{-1}(x_i/y_i)$, equation (\ref{mod06}) becomes 
\begin{eqnarray}
\label{mod06c}\dot{\theta}_i=-\omega_i, \;\; i=1,2, \ldots, N.
\end{eqnarray}
From (\ref{mod06c}) it is obvious that the angle and hence the frequency (which is similar to that of the linear harmonic oscillator) are independent of the amplitude of oscillation, irrespective of the form of $f(\bar{x})$. However, one may note that this does not always imply isochronocity as the amplitude of oscillation may decay with time, as shown in equation (\ref{sol_even}).


\begin{figure}
\begin{center}
\includegraphics[width=7cm]{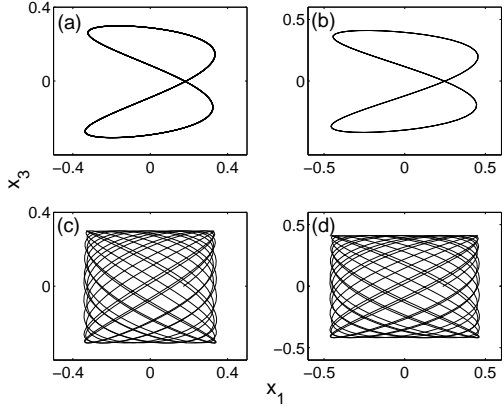}
\caption{Projected phase space of system (\ref{nhorm3a}) in the $x_1-x_3$ plane with $N=10$,  for two different values $q=3$ (Figures (a,c)) and $q=5$ (Figures (b,d)), respectively.  Figures (a) and (b) describe the $2:1$ period oscillations for the choice $\omega_1=2$ and $\omega_2,\omega_3,\ldots,\omega_{10}=1$.  Figures (c) and (d) describe the quasiperiodic oscillations for the choice $\omega_1=\sqrt{2}$ and $\omega_2,\omega_3,\ldots,\omega_{10}=1$. } 
\label{gmp2}
\end{center}
\end{figure}
\section{2$N$ integrals of motions}
\label{sec3}
In this section we show the existence of $2N$ independent integrals of motion for equation (\ref{nhorm3}) with $f=f_i$, $i=1,2,\ldots,N$, being a homogeneous polynomial (\ref{mod10e}). 
In order to show that there exists $N$ time dependent integrals, let us consider the equivalent form of (\ref{mod07}), that is, 
\begin{eqnarray}
\label{mod16}
 \frac{y_i}{x_i}\equiv\frac{(\dot{x}_i+f(\bar{x})x_i)}{\omega_ix_i}=\cot(\omega_it+\delta_i), \; i=1,2, \ldots, N.
\end{eqnarray}
Rewriting the above equation (the first and the last expressions) we get $N$ time dependent integrals as
\begin{eqnarray}
\label{mod16a}
\delta_i=\cot^{-1}\bigg[\frac{y_i}{x_i}\bigg]-\omega_it, \; i=1,2, \ldots, N,
\end{eqnarray}
where $y_i=((\dot{x}_i+f(\bar{x})x_i))/(\omega_i)$.

Now to find the remaining $N$ integrals for $q$ odd or even integer in (\ref{mod10e}) we proceed as follows.
From (\ref{mod16}) we get $\frac{x_i}{y_i}=\tan(\theta_i), \,\,\,\theta_i=\omega_it+\delta_i$.
Using this expression in the well known trignometric identity $\sin\theta_i=\frac{\tan\theta_i}{\sqrt{1+\tan^2\theta_i}}$ and $\cos\theta_i=\frac{\cot\theta_i}{\sqrt{1+\cot^2\theta_i}}$, we get
\begin{subequations}
\begin{eqnarray}
\label{mod17a}
 \sin(\omega_it+\delta_i)&=&\frac{x_i}{\sqrt{y_i^2+x_i^2}},\\
\label{mod17b}
 \cos(\omega_it+\delta_i)&=&\frac{y_i}{\sqrt{y_i^2+x_i^2}}, \; i=1,2, \ldots, N.
\end{eqnarray}
\label {mod17}
\end{subequations}
Rewriting now (\ref{mod08}) in the form $I_k=\frac{x_k}{x_N}\frac{\sin(\omega_Nt+\delta_N)}{\sin(\omega_it+\delta_i)},$ we obtain
\begin{eqnarray}
I_k^2=\frac{x^2_k}{x^2_N}\frac{\sin^2(\omega_Nt+\delta_N)}{\sin^2(\omega_it+\delta_i)}.\label{rev2}
\end{eqnarray}
Now by using (\ref{mod17a}) we can write
\begin{eqnarray}
\sin^2(\omega_it+\delta_i)=\frac{x_i^2}{y_i^2+x_i^2},\quad\sin^2(\omega_Nt+\delta_N)=\frac{x_N^2}{y_N^2+x_N^2}.
\end{eqnarray}
Substituting the above in (\ref{rev2}) we obtain the first $N-1$ time independent integrals as
\begin{eqnarray}
\label{mod15}
I_k^2&=&\frac{(x_k^2+y_k^2)}{(x_N^2+y_N^2)}, \;\;k=1,2,\ldots, N-1.
\end{eqnarray}

For $q$ odd positive integer, substituting (\ref{mod17}) and (\ref{mod15}) into (\ref{mod12}) and rearranging we arrive at the following form for the integral $I_N^o$,
\begin{eqnarray}
\label{mod14}
&&\hspace{-2cm}I_N^o=\frac{1}{(x_N^2+y_N^2)^{\frac{2m+1}{2}}}+\sum_{l=1}^N\frac{\bar{\gamma_l}}{{\omega_l}}\frac{y_l}{\sqrt{(x_l^2+y_l^2)}}\bigg(\frac{x_l^{2m-1}}{(x_l^2+y_l^2)^{\frac{2m-1}{2}}}\nonumber\\
&&\hspace{4cm}+\sum_{k=1}^{m-1}C_k^o\frac{x_l^{2m-2k-1}}{(x_l^2+y_l^2)^{\frac{2m-2k-1}{2}}}\bigg),
\end{eqnarray}
which is the $N$th time independent integral. For $m=0$ the integral (\ref{mod14}) becomes 
\begin{eqnarray}
\label{mod14a}
I_N^o&=&\frac{\sum_{l=1}^{N} \frac{\gamma_l}{\omega_l}y_l+1}{(x_N^2+y_N^2)^{\frac{1}{2}}}.
\end{eqnarray}
One can note that for $m=0$ the integrals (\ref{mod16a}), (\ref{mod15}) and (\ref{mod14a}) exactly match with those presented in \cite{Gladwin:09}. For instance, equation (\ref{mod14}) exactly reduces to the corresponding form given in \cite{Iacono:11} for the scalar case ($N=1$).   

Similarly, for the case where $q$ is even positive integer, we substitute  (\ref{mod17}) and (\ref{mod15}) into (\ref{mod12b}) and rearranging we arrive at the following form for the integral $I_N^e$,
\begin{eqnarray}
&&\hspace{-2cm}I_N^e=\frac{1}{(x_N^2+y_N^2)^{\frac{2m+1}{2}}}-Bt+\sum_{l=1}^N\frac{\bar{\gamma_l}}{{\omega_l}}\frac{y_l}{\sqrt{(x_l^2+y_l^2)}}\bigg(\frac{x_l^{2m-1}}{(x_l^2+y_l^2)^{\frac{2m-1}{2}}}\nonumber\\
&&\hspace{4cm}+\sum_{k=1}^{m-1}C_k^e\frac{x_l^{2m-2k-1}}{(x_l^2+y_l^2)^{\frac{2m-2k-1}{2}}}\bigg).
\end{eqnarray}
 We note here that the above first integral is a time dependent one.

\section{Almost integrable systems }
\label{nearlyintegrable}
In Sec. \ref{sec3} we obtained the general solution of (\ref{nhorm3}) for the choice $f_i=f$, $i=1,2,\ldots,N$ and $f$ is a homogeneous polynomial as in (\ref{mod10e}).  However, we wish to point out that the system (\ref{nhorm3}) for any arbitrary choice of (\ref{mod07e}) is almost integrable as there always exist $N$ time dependent integrals (\ref{mod16a}) and $(N-1)$ time independent integrals (\ref{mod15}).  For complete integrability only the first order nonautonomous differential equation (\ref{mod07a}) needs to be integrated.  For those forms of $f$ for which this cannot be done explicitly, one can always carry out a numerical integration of (\ref{mod07a}) or apply a suitable approximation method to find $x_N(t)$.  The remaining $x_i(t)$'s, $i=1,2,\ldots,(N-1)$, can be obtained readily using (\ref{mod08}) and $x_N(t)$.  Equation (\ref{mod06c}) ensures that if the solutions are periodic or quasiperiodic, the frequency is independent of amplitude.
\begin{figure}[!ht]
\begin{center}
\includegraphics[width=0.8\linewidth]{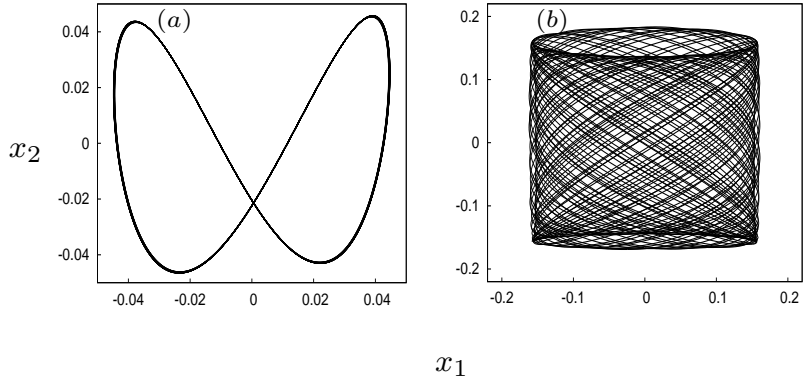}
\caption{Projected phase space of the almost integrable system (\ref{nonintegrable}) in the $x_1-x_2$ plane for the choices (a) $\omega_1=1,\omega_2=2$ exhibiting $1:2$ period oscillation, (b) $\omega_1=\sqrt{2},\omega_2=1$ exhibiting quasiperiodic oscillation} 
\label{fig4}
\end{center}
\end{figure}

In order to demonstrate the above, let us consider the special case $N=2$ in (\ref{nhorm3a}) and $f(x_1,x_2)=x_1+x_1x_2+x_2^3$.  This choice reduces equation (\ref{nhorm3a}) to the following system of coupled second order nonlinear ODEs,
\begin{subequations}
\label{nonintegrable}
\begin{eqnarray}
&&\hspace{-0.5cm}\ddot{x}_1+(x_1^2+3x_1x_2^2)\dot{x}_2+(3x_1+3x_1x_2+2x_2^3)\dot{x}_1+x_1^2x_2^2(x_1+2x_2+2x_2^2)\nonumber\\
&&\hspace{3cm}+2x_1^3x_2+x_1x_2^6+x_1^3+\omega_1^2x_1=0,\\
&&\hspace{-0.5cm}\ddot{x}_2+(3x_1+3x_1x_2+2x_2^3)\dot{x}_1+(x_1^2+3x_1x_2^2)\dot{x}_2+x_1^2x_2^2(x_1+2x_2+2x_2^2)\nonumber\\
&&\hspace{3cm}+x_1x_2^6+2x_1^3x_2+x_1^3+\omega_2^2x_2=0.
\end{eqnarray}
\end{subequations}
The solution of the above system of equations can be deduced by solving the following first order ODE obtained by the procedure discussed in the previous section,
\begin{eqnarray}
\dot{x}_1=\omega_1\cot(\omega_1t+\delta_1)x_1-x_1^2-x_1^3I_1\frac{\sin(\omega_2t+\delta_2)}{\sin(\omega_1t+\delta_1)}-x_1^4I_1^3\frac{\sin^3(\omega_2t+\delta_2)}{\sin^3(\omega_1t+\delta_1)}.\label{first}
\end{eqnarray}
However, we find that the explicit general solution of the above first order ODE is not known \cite{murphy}.  One can apply suitable numerical methods to solve this equation.  Figure \ref{fig4} is plotted by solving Eq. (\ref{first}) using a variable step size Runge-Kutta fourth order method.   We find the system (\ref{nonintegrable}) exhibits periodic and quasiperiodic oscillations which is shown in Fig \ref{fig4}.  Here the projection of the phase space of (\ref{nonintegrable}) in the $x_1-x_2$ plane for the set $(\omega_1,\omega_2)=(1,2)$ and $(\omega_1,\omega_2)=(\sqrt{2},1)$ are shown for $1:2$ periodic behaviour and quasiperiodic behaviour, respectively.  

\begin{table}
\caption{\label{table1}A comparison of the solution of Eq. (\ref{first}) obtained using  the variable step size Runge-Kutta fourth order (RK4) numerical procedure and the homotopy perturbation method (HPM).}
\begin{center}
\begin{tabular}{|l|l|l|}
\hline $t$ & $x(t)$ using variable & $x(t)$ using HPM \\
& step size RK4 &  up to third order \\
\hline
0&0.2644&0.2644\qquad\\
1&0.1556&0.1884\qquad\\
2&-0.1416&-0.1791\qquad\\
3&-0.2378&-0.2647\qquad\\
4&0.1090&0.1151\qquad\\
5&0.2381&0.2846\qquad\\
6&-0.0128&-0.0165\qquad\\
7&-0.2486&-0.2932\qquad\\
8&-0.0790&-0.0797\qquad\\
9&0.2663&0.2824\qquad\\
\hline
\end{tabular}
\end{center}
\end{table}

One can also apply perturbation techniques such as the homotopy analysis \cite{domairry} to find approximate solution of high accuracy for Eq. (\ref{first}) and compare the results with the numerical analysis. The homotopy perturbation method \cite{domairry,domairry2} involves the introduction of an artificial parameter, say $p$, into the original nonlinear equation $A(u)=0$ as
\begin{eqnarray}
(1-p)(L(u)-L(u_0))+pA(u)=0,\label{perturb}
\end{eqnarray}
where $L$ is the linear operator corresponding to the linear part of the given equation and the $A$
is the operator corresponding to the given nonlinear equation. Here $u_0$ is the lowest order approximate solution. Expressing the solution $u$ as a power series in $p$, i.e.
\begin{eqnarray}
u=u_0+pu_1+p^2u_2+p^3u_3+\ldots,
\end{eqnarray}
where $u_1$, $u_2,...,$ are the higher order approximations, one can substitute this series solution in (\ref{first}) to obtain a system of linear first order ordinary differential equations. Solving this system with suitable initial conditions one can obtain the approximate solution to the given equation in the limit $p\rightarrow 1$.  For further details on this procedure one can refer to Refs. \cite{domairry,domairry2}.  

In Table \ref{table1} we compare the solution of Eq. (\ref{first}) obtained through this procedure with the numerical solution for the parametric choice $\delta_1=1,\,\delta_2=1.5,\,\omega_1=\sqrt{2},\,\omega_2=1$.  The values listed in Table \ref{table1} for the homotopy perturbation method are calculated up to third order approximation.  We note here that the accuracy of the perturbation solution will improve if further higher order approximations are taken into the calculation \cite{domairry2}, which we do not pursue here.

\section{Conclusion}
\label{sec4}
In this paper, we have shown that a system of $N$ coupled nonlinear Lienard type oscillators admits periodic and quasiperiodic solutions or damped oscillatory periodic solutions with amplitude independent frequency of oscillations.  We have derived explicit general solution and $2N$ integrals of motion for this system. Thus we prove this system to be completely integrable.  We have also shown that another system of $N$ coupled Lienard type oscillators is almost integrable in the sense that it admits $(2N-1)$ independent integrals and reduces to a single nonautonomous first order nonlinear differential equation.  We have also shown that this almost integrable system also exhibits periodic and quasiperiodic oscillations for suitable parametric choices.  For the general system of $N$ coupled nonlinear oscillators (\ref{nhorm3}) with arbitrary form of nonlinearity, the nonlocal transformations reduce it to a system of $N$ coupled first order ODEs (\ref{mod07e}).  It will be interesting to investigate further whether other forms of nonlinearity (different from (\ref{mod10e})) are also amenable to analysis.

\section*{Acknowledgements}
The work is supported by a Department of Science and Technology (DST)--Ramanna fellowship project and a DST--IRHPA research project of M. L., who is also supported by a DAE Raja Ramanna Fellowship. JHS is supported by a DST--FAST TRACK Young Scientist research project.

\end{document}